# Experimental demonstration of broadband Lorentz non-reciprocity in an integrable photonic architecture based on Mach-Zehnder modulators

Yisu Yang,[1,*] Christophe Galland,[1,2] Yang Liu,[1] Kang Tan,[3] Ran Ding,[1] Qi Li,[4] Keren Burgman,[4] Tom Baehr-Jones[1,3] and Michael Hochberg[3,5,1]

[1]*Department of Electrical & Computer Engineering, University of Delaware, Newark DE 19716, USA*
[2]*Ecole Polytechnique Fédérale de Lausanne (EPFL), 1015 Lausanne, Switzerland*
[3]*Department of Electrical & Computer Engineering, National University of Singapore, Singapore 117576, Singapore*
[4]*Department of Electrical Engineering, Columbia University, New York, NY, USA*
[5]*Institute of Microelectronics, A*STAR (Agency for Science, Technology and Research), 11 Science Park Road, Singapore Science Park II, Singapore 117685, Singapore*
*\*yangyisu916@gmail.com*

**Abstract:** Lorentz reciprocity is a direct consequence of Maxwell equations governing the propagation of light in passive linear media with symmetric permittivity and permeability tensors. Here, we demonstrate the first active optical isolator and circulator implemented in a linear and reciprocal material platform using commercial Mach-Zehnder modulators. In a proof-of-principle experiment based on single-mode polarization-maintaining fibers, we achieve more than 12.5 dB isolation over an unprecedented 8.7 THz bandwidth at telecommunication wavelengths, with only 9.1 dB total insertion loss. Our architecture provides a practical answer to the challenge of non-reciprocal light routing in photonic integrated circuits.

## References and links

# 1. Introduction

As was recently emphasized [1,2] the realization of a true optical isolator requires breaking Lorentz reciprocity. The advent of photonic integrated circuits (PIC) has made urgent the need for non-reciprocal systems that do not rely on magneto-optic materials [3-9], which are difficult to integrate in large-scale photonic platforms [10-16]. The achievement of optical isolation without magneto-optical materials has been a long-standing challenge that has regained interest with the development of PIC in silicon and III/V material platforms. Magneto-optical materials break Lorentz reciprocity because of their asymmetric magnetic permeability and are ubiquitously employed in current bulk and fibered optical systems to achieve the key functionalities of optical isolation and circulation. Isolators are in particular necessary to protect laser sources and amplifiers from back-reflections. Circulators are devices where light injected from a first input port is transmitted to the output port, while light incoming into the output port is redirected to a third port (called the "circulated" port below). They are used to perform tasks such spectral filtering in combination with fiber Bragg gratings, and more generally they allow for versatile light routing [17]. Unfortunately, it has proved difficult to integrate magneto-optic materials in a process compatible with the scalable fabrication of monolithic PICs [10-16], leaving open the problem of how to implement isolators and circulators in these emerging platforms. This situation has triggered intense research efforts toward alternatives to magneto-optic materials, one of them being the use of optical nonlinearities [7-9] that can be enhanced in nanoscale cavities such as ring resonators. However, this approach is intrinsically narrow-band (due to the resonance condition) and the performance is power-dependent, limiting the widespread applicability of such devices. An overview of some recent results obtained with the above-discussed techniques is presented in the Appendix, Table 2 (the reader is referred to [18] for a more comprehensive review).

Probably the most promising route toward a practical non-reciprocal system entirely compatible with existing PIC technologies is to use a time-dependent modulation of the refractive index to break time-reversal symmetry, which can be interpreted as a photonic Aharonov-Bohm effect [19-21]. Following the pioneering work of Noé and coworkers [3], a few recent experimental results [4-6] have shown promise along this line of research, but still suffer from either low isolation, prohibitively high insertion loss, or narrow-band operation, and can be complex to implement.

# 2. Isolator and circulator system design

In this paper, we introduce a new architecture for a time-modulated optoelectronic system that performs the key functions of optical isolator and circulator with a minimum level of complexity. Although, at present, the system is realized with discrete optoelectronic devices, it is a first proof of principle of a design that is compatible with existing PIC technologies such as integrated silicon photonics. In previous approaches based on traveling-wave modulators [3,5] reciprocity could only be broken if the light travel time through the modulators was larger than the period of the modulation signal, requiring both long modulators and high (>1 GHz) driving frequencies. We overcome these requirements by using a two-stage architecture in which two Mach-Zehnder modulators are separated by a passive optical delay line [6,22]. This strongly relaxes the demand on modulator speed and geometry, as the modulation frequency is inversely proportional to the light propagation time inside the delay line instead of inside the modulator itself. Compared to the scheme used by Doerr *et al.* [6], we achieve efficient and broadband isolation over the telecommunication band with only two modulators, without the need for a multiplicity of active systems in parallel. The main characteristics and performances of our system are compared to previous results on modulation-based isolators in Table 1. Last, but not least, we are the first to demonstrate a device that is also an optical *circulator* – a vital component in large-scale photonic circuits – without relying on magneto-optical materials [23,24].

| Reference | Traveling-wave modulators | Delay line | Modulation signal | Circulator | Bandwidth (THz) | Insertion loss (dB) | Extinction Ratio (dB) |
|---|---|---|---|---|---|---|---|
| Lira *et al.* [5] | Yes | No | Sine, 10 GHz | No | 0.2 | 70 | 3 |
| Doerr *et al.* [6] | No | Yes | Sine, 2 GHz | No | 5 | 11 | 3 |
| Bhandare *et al.* [3] | Yes | No | Sine, 2 GHz | No | 3.7 | 23.8 | 30 |
| Li *et al.*[21](off-chip) | Yes | No | Sine, 50MHz | No | One VIS wavelength | NA | 13 |
| This work *(off-chip proof of principle)* | No | Yes | Square, 20 MHz | Yes | 8.7 | 9.1 | 12.5 |

**Table 1**. Summary of time-modulated optical isolator systems that achieve broadband isolation

By making use of a non-sinusoidal drive signal, a time-modulated optical system can be constructed that violates reciprocity in what we conjecture is the simplest possible architecture for broadband isolation. With a square-wave signal, perfect isolation can be achieved in theory. Moreover, by using a passive delay line, the modulation speed can be lowered to 20 MHz instead of several GHz. Fig. 1(a) shows a block diagram of the system. Due to the very low speeds involved, we note that the modulators are expected to function essentially as lumped elements.

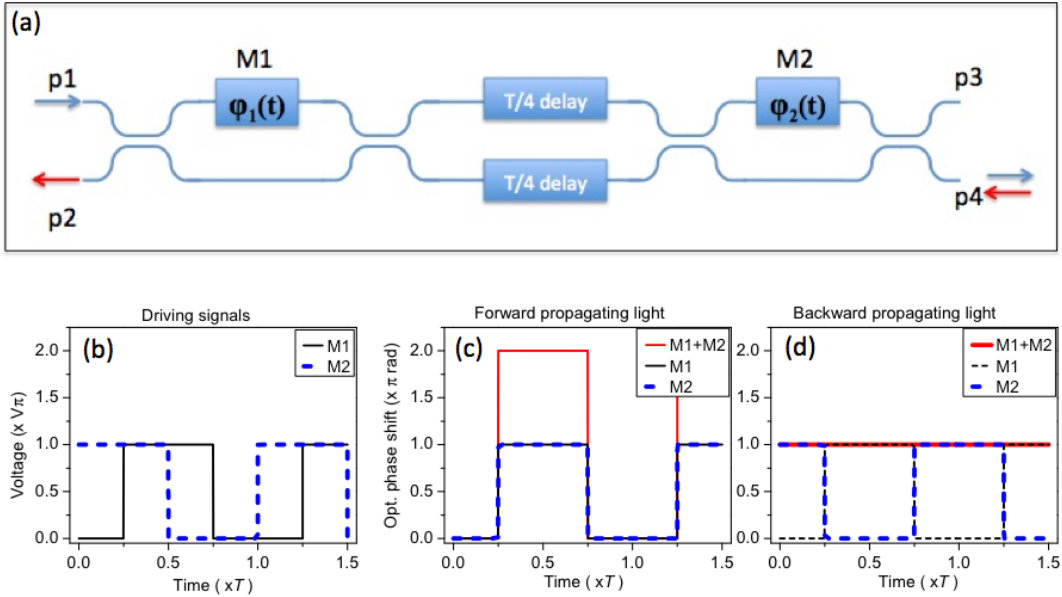

**Fig. 1.** Schematics and working principle of the optical isolator/circulator system. (**a**), Block diagram of our non-reciprocal system, consisting of two Mach-Zehnder modulators in series, separated by a fiber-based delay line. The path for forward propagation of light is shown in blue, from port p1 to port p4. In contrast, the reverse flow of light is redirected in a nonreciprocal manner from port p4 to p2, as shown by the red arrows. (**b**), Schematic time-domain square-wave signals used to drive the two modulators, in units of Vπ, the voltage corresponding to a π optical phase shift. The time axis is in units of the period $T$ of the drive signals. (**c**), Computed optical phase modulation experienced by the forward and (**d**) reverse propagating modes, evidencing how our scheme breaks Lorentz reciprocity.

As shown in Fig. 1(a), the signals driving the two modulators are shifted by a time lag matching the propagation time in the optical delay line, which is equal to one quarter of the modulation period. As a result, when each modulator is driven with an amplitude Vπ (the voltage required to apply a π optical phase shift, Fig. 1(b)) the forward-propagating mode experiences an accumulated optical phase shift of 0 or 2π radians (Fig. 1(c)), while the reverse propagating mode sees a net optical phase shift of π radians (Fig. 1(d)). The offset bias of the system has been chosen so that the forward propagating mode is coupled from p1 to p4, while the reverse mode couples from p4 to p2. Thus, non-reciprocal coupling and both isolator and circulator functionalities are achieved. A detailed analytical derivation of the port-to-port transmission of the system is presented in Section 2 of the Appendix. We note that if an ideal square wave is provided, perfect isolation is expected. In practice, the condition for isolation is not satisfied at the rising and falling edges of the square wave if these are not perfectly sharp. As long as the modulation frequency $f=1/T$ is small compared to the bandwidth of the modulators, the duration over which the circulator function is impaired will be relatively small as shown in the Appendix. This condition is well satisfied in our system, with a measured rise and fall time of 4 ns (see Fig. 8), corresponding to a bandwidth of 250 MHz. It is therefore advantageous to use a long delay line for optimal isolation, while a trade-off with increasing insertion loss must be considered.

## 3. Experiment results and discussion

Our isolator/circulator system was realized in polarization-maintaining (PM) single-mode optical fibers by means of two EOSpace X-cut Lithium Niobate 2x2 modulators configured for push-pull operation. For simplicity and without loss of generality, we replaced push-pull configurations in the schematics by the equivalent system as shown in Fig. 1(a). Detailed analysis is presented in the Appendix, Section 2. The $V\pi$ was approximately 4.6 V. In addition to the modulation signal, a d.c. offset on each modulator was applied to properly bias the phase imbalance of the interferometers. We also used two General Photonics manual delay lines to balance the path lengths between the two 2.5 m long arms of the optical delay line to an accuracy of better than 100 ps, a key requirement for broadband performance. An arbitrary function generator (AFG, Tektronix AFG3252C) provides the modulators' RF drive voltage with 4.5 V peak-to-peak amplitude. Given the 2.5 m long optical path of the delay line the corresponding time-shift of the two driving signals is $T/4 = 12.4$ ns, corresponding to an optimal modulation frequency $f = 20.2$ MHz.

All components in the system were polarization-maintaining, and the laser was linearly polarized. Thus, only a single optical mode exists, common to all ports, which is an essential requirement to prove unambiguously the non-reciprocal nature of our system [1,2]. The delay lines were adjusted to give as close to a balanced path as possible, which was correlated with the optical bandwidth seen in a spectral sweep of the circulator transmission from p1 to p4. In the course of our experiments as shown in Fig. 2, the d.c. offset on the modulators usually had to be adjusted slightly between measurements every 10 minutes on average. This is due to the fluctuating phase imbalance between the optical paths in the delay line and could easily be mitigated by an active feedback stabilization loop, which would however add unnecessary complexity to our proof-of-principle experiment. Moreover, it is expected that an integrated version of our architecture would be intrinsically much more stable. The delay lines and modulation frequency were not adjusted within each set of steady state and time-domain measurements shown below, but adjustment was performed between them. In particular, the frequency was adjusted by around 2%. This is likely due to a combination of fluctuations in effective path lengths and possible instabilities in the arbitrary function generator's internal clock.

To balance the two optical paths of the delay line we first maximize the Free Spectral Range (FSR) of the system's transmission without modulation. Accurate balance is essential for optimal optical bandwidth. In forward configuration, when 0 dBm of continuous wave (CW) laser beam was sent to port p1, port p4 output was measured. In reverse configuration, keeping all system's parameters unchanged, we sent the same laser beam into port p4 and measured port p1's output power. The laser was swept from 1470 to 1570 nm. Overall power conservation in forward vs. reverse propagation was verified by measuring the ports p2 and p3 outputs.

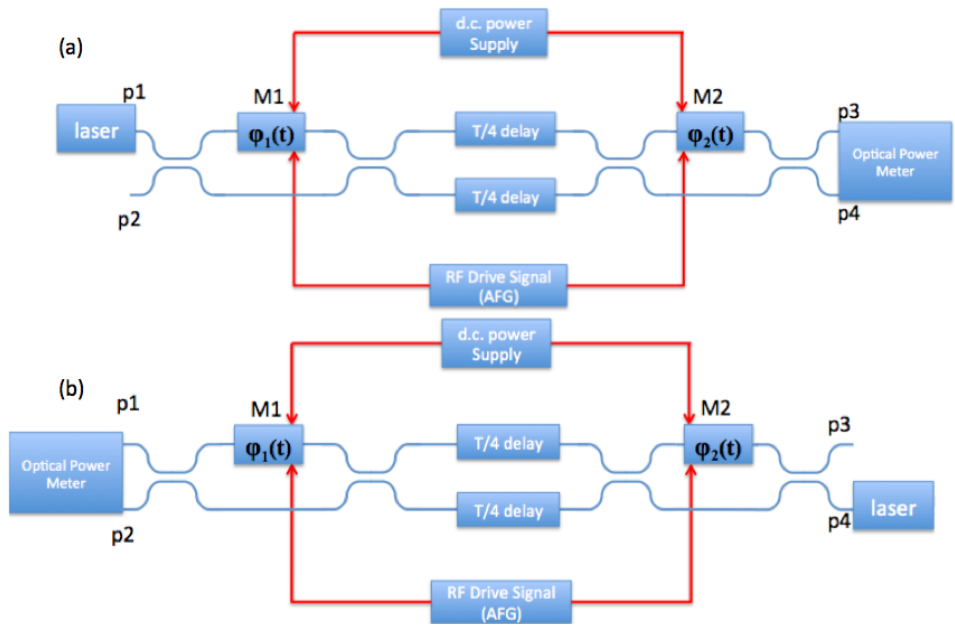

**Fig. 2**. Schematic of the setup used in the experiments: $M_1$, $M_2$: Mach-Zehnder Modulators 1 and 2; AFG: arbitrary function generator; red line: electrical signal path, blue line: optical path. (**a**) Forward transmission measurement (**b**) Reverse transmission measurement

The main results of isolation are presented in Fig. 3. The maximum isolator/circulator system excess loss is 9.2 dB with an isolation ratio of 12.5 dB or more across the wavelength range 1500 – 1568 nm, which corresponds to a record optical bandwidth of 8.7 THz. The total loss during active operation was only 1 dB more than when the modulators were biased at full transmission and the modulation signal was powered off.

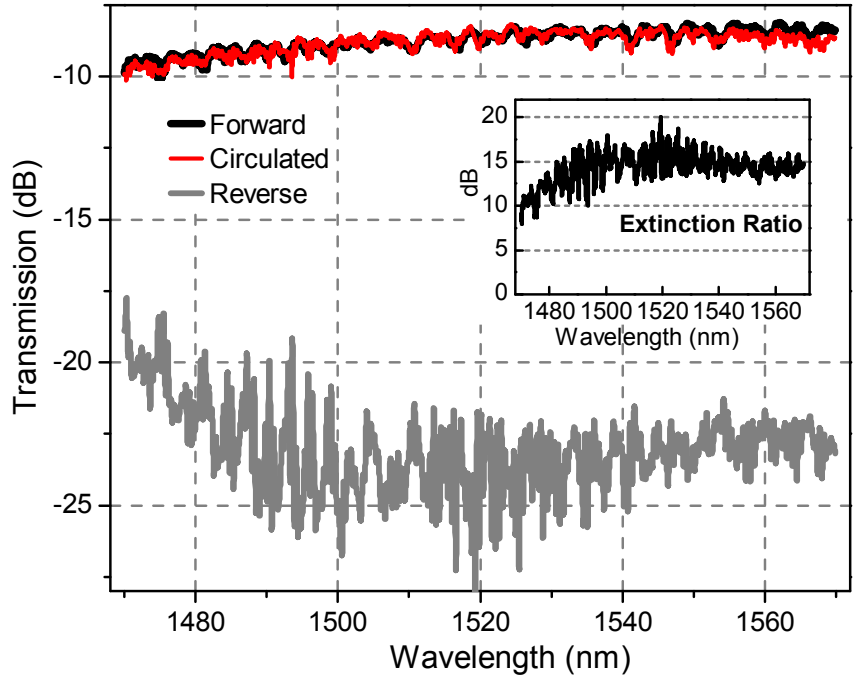

**Fig. 3.** Experimental characterization of the isolator/circulator system. The laser wavelength was swept from 1470 nm to 1570 nm while measuring the transmission through the forward path from p1 to p4 (black line), the reverse path from p4 to p1 (grey line), and the circulated path from p4 to p2 (red line). The extinction ratio of the isolator (the difference between the black and grey curves) is plotted in the inset. It reaches close to 20 dB at some wavelengths and is over 12.5 dB over the 1500 – 1568 nm window.

The real-time output powers of the system in forward and reverse configurations were measured as shown in Fig. 4. Time domain characterization of the system transmission was performed with a 1.4 GHz bandwidth avalanche photodetector (APD) and an optical power meter. We could obtain the average power sent to the APD while recording the real-time output signal with a fast oscilloscope (Agilent DSO7014A 100 MHz 2GSa/s). A CW laser beam (wavelength = 1555.51 nm, power = -8 dBm) was sent successively through the path p1 to p4 (forward path), through p4 to p1 (isolated path), and p4 to p2 (circulated path) to observe any possible fluctuations in the time domain transmission. Experimental results showed that typical amplitudes fluctuations were lower than 1.0% for the forward path, while fluctuations of 23.4% were seen for the isolated path and 1.6% for the circulated path as shown in Fig. 6(a). The fluctuations on the isolated path are low in absolute terms, given the high extinction on this port (Inset in Fig. 6(a)). The isolation performance is weakened by the non-ideal square-wave with a rise and fall time of 4 ns as shown in the Appendix, Section 2. But the averaged isolation ratio is still better than 10 dB.

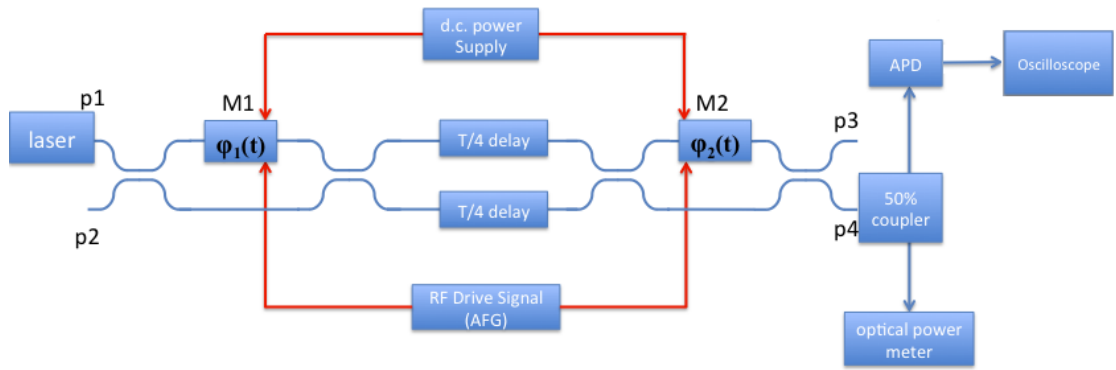

**Fig. 4**. Time domain characterization of the system transmission (Forward configuration is shown. Red line: electrical signal path. Blue line: optical path)

In view of the important applications of PICs in high-speed digital telecommunications, we characterized the impact of our system on the data transmission fidelity. Due to the system's working principle, the optical phase is flipped twice per period, which could be problematic if a phase-modulated signal format was sent to the system. Therefore we limit our study here to OOK modulation.

The testing setup is shown in Fig. 5. We sent an optical signal encoding a 25 Gb/s non-return-to-zero (NRZ) pseudo-random bit sequence (PRBS was generated by Tektronix Arbitrary Waveform Generator AWG70001A) through the circulator from port p1 to p4, and measured the output on a sampling scope (Tektronix DSA8200 Digital Serial Analyzer, 10 GHz bandwidth). Compared to the signal without the circulator (Fig. 6(b)), some additional dispersion in the level of the "1" bit was observed, but the so-called "eye" of the diagram remained well open (Fig. 6(c)), showing that faithful digital data transmission is not impaired by the isolator/circulator.

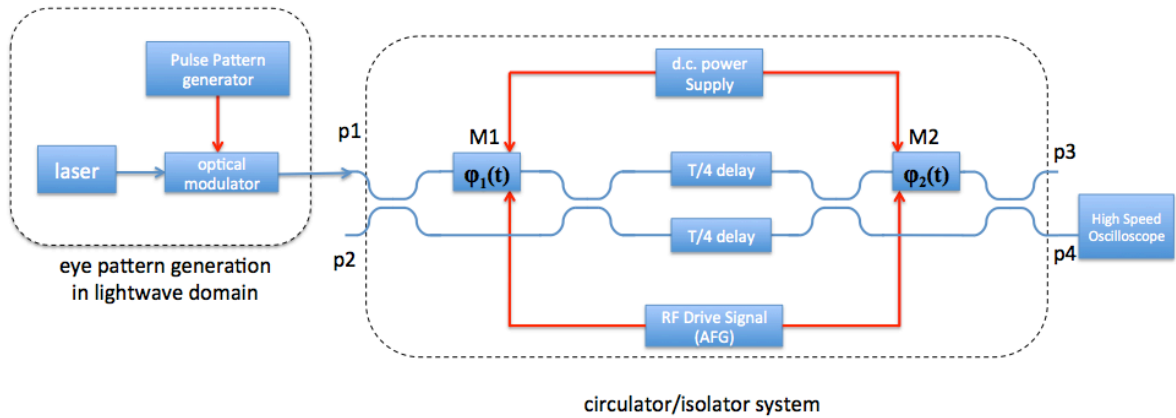

**Fig. 5**. Schematic of the data transmission fidelity measurement setup. Red line: electrical signal path. Blue line: optical path

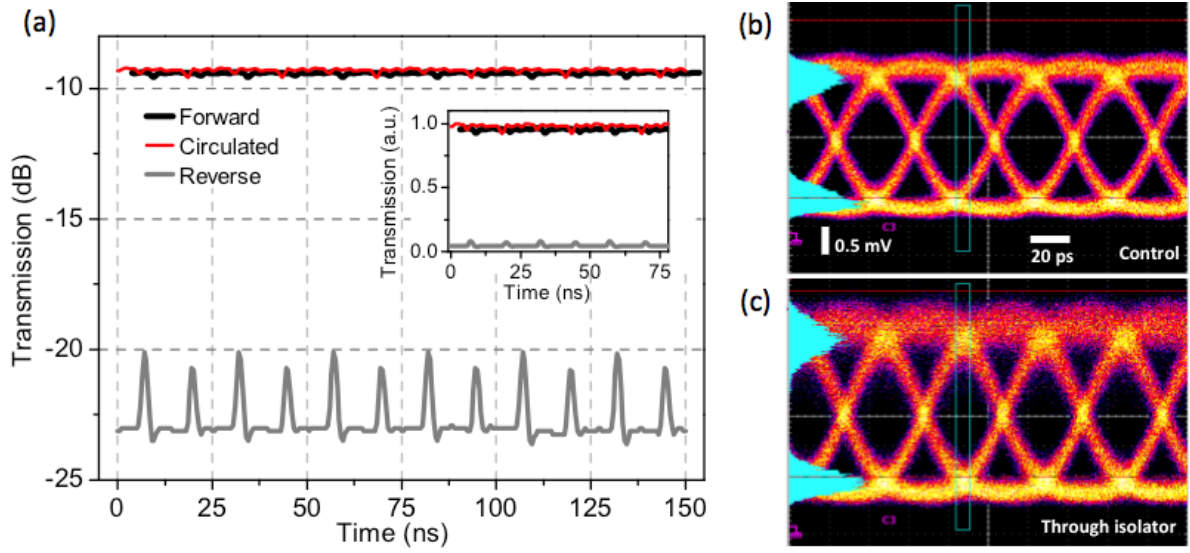

**Fig. 6.** Time-domain characterization of the circulator. **(a)**, For a CW input light, transmission from p1 to p4, forward path (black line), as well as transmission from p4 to p1 (isolated path, grey line) and p4 to p2 (circulated path, red line). In the inset the data is plotted on a linear, normalized vertical scale, to show the absolute magnitude of the residual intensity modulations on the transmitted light. **(b)**, Pseudo-random bit sequence (PRBS) pattern measured on a sampling scope before and **(c)** after transmission through the circulator system (path p1 to p4).

One key question is whether the circulator system presented here could indeed be implemented in integrated silicon photonics. Gathering some of the most recent results from the literature [25-32], we predict that a 1.5 mm long modulator could provide the needed π phase shift (if both arms are driven) with only 3.5 dB of loss [28], while the needed delay line could be implemented with 1.2 µm wide silicon ridge waveguides [29]. In such waveguides, group indices of around 4.3 and losses of 0.3 dB/cm are obtained. Therefore, a 4 cm long on-chip delay line, which entails only 1.2 dB of loss, could enable a modulation frequency around 400 MHz. At such a low speed, a high-quality square wave can be generated and imparted onto the optical phase, considering the reported silicon modulator speeds of 30 GHz [30-32]. Due to group velocity dispersion, the optical delay between M1 and M2 is actually wavelength dependent. Consider a 500 x 220nm$^2$ silicon wire waveguide surrounded by SiO$_2$ as the delay line, the group dispersion is about $1.2 \times 10^{-4}$ /nm around 1550nm. We estimate a delay variation of less than 1ps over more than 100 nm bandwidth around 1550 nm [22]. This is much smaller than the modulation period (2500 ps here) and has therefore negligible impact on performance. The total insertion loss of this system would be 8.2 dB, while its extinction ratio is expected to exceed the values reported here thanks to the intrinsic stability of integrated Mach-Zehnder interferometers compared to fiber optics.

## 4. Conclusions

Based on the concept introduced in Ref. [22], we have presented a simple architecture to break Lorentz reciprocity using neither magneto-optical effects nor non-linear effects but relying on commercial optical modulators. We have built a fiber-based optical circulator that achieves a level of performance approaching the needs of practical applications. Our approach can be adapted to silicon photonic circuits and other integrated optics platform with equal or better performance assuming existing state of the art technologies. Our design is a practical approach to optical isolation in future integrated photonics systems and can become an enabling component for further developments in this field.

**Acknowledgments**

C. G. would like to acknowledge the financial support of the Swiss National Science Foundation through an *Ambizione* Fellowship. All authors gratefully acknowledge support from an AFOSR STTR grant, number FA9550-12-C-0079 and FA9550-12-C-0038. The authors would like to thank Gernot Pomrenke, of AFOSR, for his support of the OpSIS effort, though both a PECASE award (FA9550-13-1-0027) and ongoing funding for OpSIS (FA9550-10-1-0439), Brett Pokines and AFOSR SOARD office, for their support of the development of high-speed modulators under grant FA9550-13-1-0176. The authors are grateful for support from an MOE ACRF Tier-1 NUS startup grant and NRF NRF2012NRF-NRFF001-143. The authors would like to gratefully acknowledge AT&T for the loan of critical equipment.

# 5. Appendix

## 5.1. Overview of on-chip isolators

To the best of our knowledge, reported on-chip isolators with state-of-the-art performance are highlighted in Table 2, together with their working principles.

Table 2. Summary of recent integrated optical isolators.

| Reference | Material | *ER (dB) | Bandwidth (nm) | Insertion Loss (dB) | Other features |
|---|---|---|---|---|---|
| T. Shintaku [33] | garnet | 27 | NA | 2~5 | magneto-optic waveguide |
| J.Hwang et al. [34] | liquid crystals | 11 | 50 | 1 | electro-tunable optical diode; working wavelength is around 500nm |
| H.Shimizu et al. [35] | InGaAsP,InP,Fe | 14.7 dB/mm | 30 | 14.1dB/mm | InGaAsP Semiconductor Optical Amplifier (SOA) |
| W.V. Parys, et al. [36] | InGaAsP,InP,Co50Fe50 | 99 dB/cm | 1 | 18 | InGaAsP SOA |
| H. Yokoi et al. [37] | Si,SiO2 and garnet | 21 | 35 | 8 | bonding garnet layer and silicon MZI; on-chip device length=4mm |
| L.Bi et al. [12] | Si,SiO2 and garnet | 19.5 | 1.6 GHz | 18.8 | on-chip device length=290μm; bonding silicon chip and garnet film |
| M. Tien et al. [13] | Si,SiO2 and garnet | 9 | 0.1 | NA | on-chip device; bonding garnet (Ce:YIG) onto a silicon ring resonator with diameter=1.8mm |
| S. Ghosh et al. [16] | Si,SiO2 and garnet | 25 | 0.5 | 8 | adhesive bonding; nonreciprocal phase shift=0.29$\pi$ |
| Y. Shoji et al. [14] | Si,SiO2 and Garnet | 30 | 5 | 13 | chip size is 1.5mm$^2$; bonding technology |
| Li Fan et al. [8] | Si and SiO2 | 28 | 1 | 12 | silicon ring resonators |
| Li Fan et al. [9] | Si and SiO2 | 40 | 0.1 | 15.5 | silicon ring resonators; 2.3mW input power |
| S. Manipatruni et al. [38] | Si and SiO2 | 20 | 0.25 | 0.1 | on-chip device using Air/Si/SiO2 DBR with mirror size: 100 μm2 |

magneto-opitcal Kerr effects

nonlinearity effects

MEMS tunning effects

*ER=extinction ratio

## 5.2. Theoretical analyzing of the isolator/circulator system

In this section, we drive analytical expressions for the port-to-port, time-dependent transmission coefficients of our system, using the transfer matrix formalism. The complex electric field amplitudes at the input and the output of a four-port device as shown in Fig. 1(a) can be written in vector notation

$$\begin{pmatrix} p_1 \\ p_2 \end{pmatrix}, \begin{pmatrix} p_3 \\ p_4 \end{pmatrix}. \tag{1}$$

For a clearer exposition, we first neglect the loss of each component in Fig. 1(a). Below a more realistic model will be considerer including the loss, the non-idealities in the modulators and the imbalance of optical paths in the delay line. The scattering matrix of an ideal 50/50 directional coupler (DC) writes

$$M_{DC} = \frac{1}{\sqrt{2}} \begin{pmatrix} 1 & i \\ i & 1 \end{pmatrix}. \tag{2}$$

The scattering matrix of phase modulators shown in Fig. 1(a) can be written

$$M_1 = \begin{pmatrix} \exp[i\phi_1(t)] & 0 \\ 0 & 1 \end{pmatrix}, M_2 = \begin{pmatrix} \exp[i\phi_2(t)] & 0 \\ 0 & 1 \end{pmatrix}. \tag{3}$$

where the optical phase shifts ($\phi_1(t)$, $\phi_2(t)$) are created by modulators $M_1$ and $M_2$. They are time dependent as shown in Fig. 1(b). The scattering matrix corresponding to the two optical delay lines between $M_1$ and $M_2$ writes

$$M_{Delay} = \begin{pmatrix} i & 0 \\ 0 & i \end{pmatrix}. \tag{4}$$

Therefore the total scattering matrix of the system in forward propagation is

$$M_{sys-fwd} = M_{DC} \times M_2 \times M_{DC} \times M_{Delay} \times M_{DC} \times M_1 \times M_{DC}. \tag{5.1}$$

$$M_{sys-fwd} = \frac{1}{4} \begin{pmatrix} 1 & i \\ i & 1 \end{pmatrix} \begin{pmatrix} e^{i\phi_2\left(t+\frac{T}{4}\right)} & 0 \\ 0 & 1 \end{pmatrix} \begin{pmatrix} 1 & i \\ i & 1 \end{pmatrix} \begin{pmatrix} i & 0 \\ 0 & i \end{pmatrix} \begin{pmatrix} 1 & i \\ i & 1 \end{pmatrix} \begin{pmatrix} e^{i\phi_1(t)} & 0 \\ 0 & 1 \end{pmatrix} \begin{pmatrix} 1 & i \\ i & 1 \end{pmatrix}. \tag{5.2}$$

$$M_{sys-fwd} = \frac{1}{2} \begin{pmatrix} -i\left[e^{i\phi_1(t)} + e^{i\phi_2\left(t+\frac{T}{4}\right)}\right] & e^{i\phi_1(t)} - e^{i\phi_2\left(t+\frac{T}{4}\right)} \\ -e^{i\phi_1(t)} + e^{i\phi_2\left(t+\frac{T}{4}\right)} & -i\left[e^{i\phi_1(t)} + e^{i\phi_2\left(t+\frac{T}{4}\right)}\right] \end{pmatrix}. \qquad (5.3)$$

For light sent into port $p_1$, the input vector is

$$\begin{pmatrix} p_1 \\ p_2 \end{pmatrix} = \begin{pmatrix} 1 \\ 0 \end{pmatrix}. \qquad (6)$$

The output signal at $p_3$ and $p_4$ is

$$\begin{pmatrix} p_3 \\ p_4 \end{pmatrix} = M_{sys} \begin{pmatrix} p_1 \\ p_2 \end{pmatrix} = \begin{cases} \begin{bmatrix} 0 \\ -1 \end{bmatrix} t \in (0, T/4), (3T/4, T] \\ \begin{bmatrix} 0 \\ 1 \end{bmatrix} t \in [T/4, 3T/4] \end{cases}. \qquad (7)$$

Eq. (7) shows that all the light sent from $p_1$ is directed to port $p_4$, with only an imprinted phase modulation. The system's transfer matrix in reverse operation is

$$M_{sys-rev} = M_{DC} \times M_1 \times M_{DC} \times M_{Delay} \times M_{DC} \times M_2 \times M_{DC}. \qquad (8.1)$$

$$M_{sys-rev} = \frac{1}{2} \begin{pmatrix} -i\left[e^{i\phi_1\left(t+\frac{T}{4}\right)} + e^{i\phi_2(t)}\right] & e^{i\phi_2(t)} - e^{i\phi_1\left(t+\frac{T}{4}\right)} \\ e^{i\phi_1\left(t+\frac{T}{4}\right)} - e^{i\phi_2(t)} & -i\left[e^{i\phi_1\left(t+\frac{T}{4}\right)} + e^{i\phi_2(t)}\right] \end{pmatrix}. \qquad (8.2)$$

The complex amplitude of input signals to $M_2$'s $p_4$ is

$$\begin{pmatrix} p_3 \\ p_4 \end{pmatrix} = \begin{pmatrix} 0 \\ 1 \end{pmatrix}. \qquad (9)$$

Referring Fig. 1(d), the output signal of $M_1$'s $p_1$ and $p_2$ should be

$$\begin{pmatrix} p_1 \\ p_2 \end{pmatrix} = M_{sys-rev} \begin{pmatrix} p_3 \\ p_4 \end{pmatrix} = \begin{cases} \begin{bmatrix} 0 \\ i \end{bmatrix} & t \in (0, T/2) \\ \begin{bmatrix} 0 \\ -i \end{bmatrix} & t \in [T/2, T] \end{cases}. \quad (10)$$

Therefore, in reverse direction, light sent from $p_4$ will only come out at $p_2$ and Lorentz reciprocity is broken.

As we take into consideration of the real waveform of the driving signals, the optical losses, the optical path lengths difference in the delay line, and d.c. bias drift from the 3 dB point in the Mach-Zehnder modulators, the model can be refined for real experimental conditions. The directional coupler's scattering matrix is modified to

$$M_{DC}(r,k) = 10^{-k/20} \begin{pmatrix} \sqrt{1-r} & i\sqrt{r} \\ i\sqrt{r} & \sqrt{1-r} \end{pmatrix}. \quad (11)$$

where $r$ is the power coupling ratio (ideally 0.5) and $k$ is the insertion loss (unit: dB). The transfer matrix of the phase modulators can be modified to

$$M_1(t,k_1) = \begin{pmatrix} 10^{-k_1/20} \exp[i\phi_1(t)] & 0 \\ 0 & 10^{-k_1/20} \end{pmatrix}, M_2(t,k_2) = \begin{pmatrix} 10^{-k_2/20} \exp[i\phi_2(t)] & 0 \\ 0 & 10^{-k_2/20} \end{pmatrix}. \quad (12)$$

Let $M_1$ insertion loss = 3.4 dB and $M_2$ insertion loss = 3.2 dB. Let $k_i$ ($i$=1,2) to be the insertion loss (unit: dB) of $M_i$. The optical delay lines' transfer matrix is modified as below:

$$M_{Delay}(k_{top}, k_{bot}) = \begin{pmatrix} 10^{-k_{top}/20} \exp(i\phi_{top}) & 0 \\ 0 & 10^{-k_{bot}/20} \exp(i\phi_{bot}) \end{pmatrix}. \quad (13)$$

The drive voltage signal is simulated by a combination of square wave, raised-cosine wave and decayed sine wave as shown below,

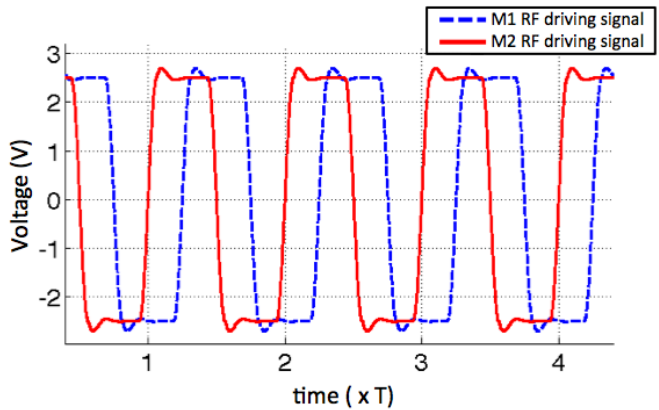

**Fig. 7**. Simulation of the input signals driving the modulators $M_1$ and $M_2$.

The decayed sine pulse that overlaps on the square wave is

$$A_1 e^{-A_2 \frac{t}{T}} sin(2\pi f_1 t). \qquad (14)$$

Where $A_1 = 0.4$, $A_2 = 16$, $T$ is the period of RF driving signals, $f_1=5*f_{RF\_drive}$. These parameters are obtained from the experimental results as shown below.

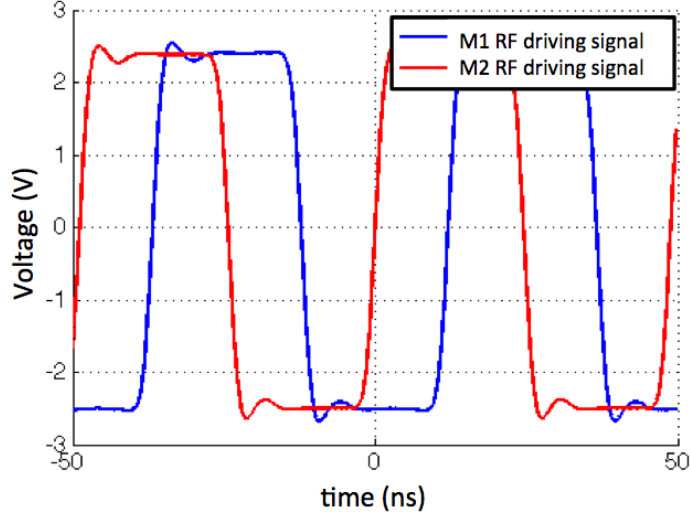

**Fig. 8**. Experimental measurement of the signal driving the modulators

A raised-cosine function is adopted to connect the high level and low level of the square wave in order to simulate the real signal. Based on Fig. 5, the rise and fall time is about 4 ns, so the raised cosine factor ($\beta$) is 0.25 in the raised-cosine function as shown below.

$$\frac{V_{pp}}{2}\left(1+cos\left[\frac{\pi f_{sa}}{\beta}\left(-t-\frac{1-\beta}{2f_{sa}}\right)\right]\right)-\frac{V_{pp}}{2}. \qquad (15)$$

where $f_{sa}=2*f_{RF\_drive}$, $V_{pp}$ is the RF driving signals' peak-to-peak voltage. The real-time simulation of the port-to-port transmission is shown in Fig. 6. Simulation parameters are summarized in the Table 3 and the simulated time averaged extinction ratio is about 23.7 dB with insertion loss of 9.5 dB. Because the losses of the components (e.g. DCs and delay lines) in the system's top and bottom arms are different, as listed in Table 3, the successive peaks and dips of real time output signal are not identical, as shown in Fig. 9. This simulation is verified by our experimental results as shown in Fig. 6(a). Our simulation also proves that when the first modulator's driving voltage is not exactly T/4 delayed compared to the second one the ER remains higher than 18dB for a time delay deviation from T/4 less than T/10.

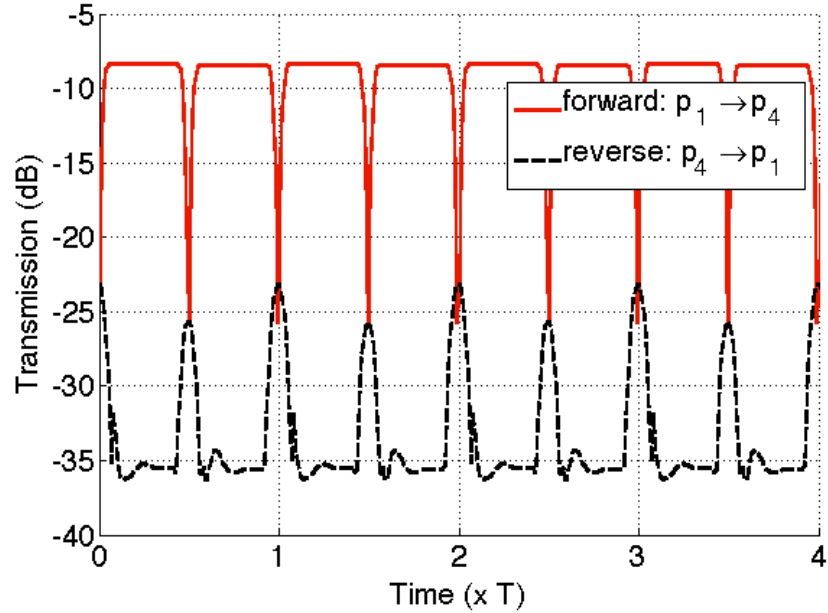

**Fig. 9**. Simulated transmission of our circulator system in forward and reverse direction

**Table 3**. Simulation parameters of the isolator/circulator system

| Parameters | Value | Unit |
| --- | --- | --- |
| $M_1$ $V\pi$ | 4.6 | V |
| $M_2$ $V\pi$ | 4.6 | V |
| $Vpp$ of drive signal | 4.5 | V |
| Splitter coupling ratios | 0.5,0.55,0.49,0.5 | NA |
| Splitter losses | 0.1,0.1,0.2,0.2 | dB |
| Loss of upper optical delay path | 1.2 | dB |
| Loss of lower optical delay path | 1.1 | dB |
| Optical path length between $M_1$ and $M_2$ | 2.51 | m |
| Length difference between bottom path and top path | 0.1 | mm |
| $M_1$ insertion loss | 3.4 | dB |
| $M_2$ insertion loss | 3.2 | dB |
| Group index of light in fiber | 1.48 | |
| Modulation frequency | 20 | MHz |
| Raising cosine factor | 0.25 | |